\documentclass[12pt]{iopart}
\usepackage{iopams}  
\usepackage{epsfig}  
\bibstyle{plain}
\begin{document}
 \title{Noise-guided evolution within cyclical interactions}
 \author{Matja{\v z} Perc$^\star$ and Attila Szolnoki$^\dagger$}
 \address{$^\star$Department of Physics, Faculty of Natural Sciences and Mathematics, University of Maribor, Koro{\v s}ka cesta 160, SI-2000 Maribor, Slovenia\\ $^\dagger$Research Institute for Technical Physics and Materials Science, P.O. Box 49, H-1525 Budapest, Hungary}
 \ead{matjaz.perc@uni-mb.si, szolnoki@mfa.kfki.hu}
 \begin{abstract}  We study a stochastic predator-prey model on a square lattice, where each of the six species 
has two superior and two inferior partners. The invasion probabilities between
species depend on the predator-prey pair and are supplemented by Gaussian
noise. Conditions are identified that warrant the largest impact of noise on
the evolutionary process, and the results of Monte Carlo simulations are
qualitatively reproduced by a four-point cluster dynamical mean-field
approximation. The observed noise-guided evolution is deeply routed in
short-range spatial correlations, which is supported by simulations on other
host lattice topologies. Our findings are conceptually related to the
coherence resonance phenomenon in dynamical systems via the mechanism of
threshold duality. We also show that the introduced concept of noise-guided
evolution via the exploitation of threshold duality is not limited to
predator-prey cyclical interactions, but may apply to models of evolutionary
game theory as well, thus indicating its applicability in several different fields of research.
 \end{abstract}
 \maketitle
 \section{Introduction}
 Cyclical interactions, the simplest non-trivial example being the children's rock-scissors-paper game, are despite their apparent simplicity fascinating examples of evolutionary processes \cite{hofbauer_88,nowak_06}. Examples from real life are the mating strategies of side-blotched lizards \cite{sinervo_n96}, overgrowths by marine sessile organisms \cite{burrows_mep98}, and competitions among different strains of bacteriocin-producing bacteria \cite{szabo_pre01a}. Moreover, studies of cyclical interactions may provide insights into the formation of defensive alliances \cite{szabo_pre01b,szabo_jpa05} and Darwinian selection \cite{maynard_n73}, as well as structural complexity and pre-biotic evolution \cite{rasmussen_s04}. Thus, cyclical interactions are rightfully acquiring a central role in the study of evolutionary processes \cite{kerr_n02}, not just in Lotka-Volterra like predator-prey systems, but also in evolutionary game theory \cite{weibull_95}, where strategic complexity \cite{hauert_s02,szabo_prl02} often leads to a closed loop of dominance between participating strategies.

An interesting concept that has only recently begun to supplement models of evolutionary processes is the addition of stochasticity at some level of interactions. While the origin of stochasticity can sometimes be related to the finiteness in population size \cite{shnerb_pnas00,traulsen_prl05}, uncertainties may also enter under the assumption of irrationality and errors in decision-making \cite{blume_geb93, szabo_pre98,traulsen_pre06c}. Moreover, the approach is viable when stochasticity in payoffs originates from the heterogeneity of the system \cite{traulsen_jtb07a}, or if explicit payoff fluctuations are considered \cite{perc_njp06a}. The latter examples are closely related to the concept we adopt presently, as stochasticity does not originate from finite populations but is due to uncertainties arising by predator-prey interactions. Irrespective of these particularities, however, stochasticity appears to posses the ability of having a profound impact on the evolutionary process \cite{traulsen_pre06b}. Examples range from stochastic gain in population dynamics \cite{traulsen_prl04} and eradication of coexistence in the cyclic Lotka-Volterra model \cite{reichenbach_pre06}, to cooperation promotion in the spatial prisoner's dilemma game \cite{perc_njp06b}.    

Already long before making its debut by evolutionary models, stochasticity has been identified as a potentially crucial agonist in certain types of dynamical systems. The so-called stochastic resonance, standing for the resonant noisy enhancement of the correlation between the system's response and a weak external stimulus, has been first reported for bistable systems \cite{benzi_jpa81}, and latter on also for a broad variety of other physical as well as biological systems \cite{gammaitoni_rmp98}. Importantly, noise can play an ordering role even in the absence of additional external signals, whereby the established term describing the phenomenon is coherence resonance \cite{hu_prl93,rappel_pre94,pikovsky_prl97,han_prl99,perc_njp05,ushakov_prl05}. Over the years a particular property of dynamical systems, termed excitability, has crystallized as being very beneficial when noise-induced phenomena are a desirable feature. Excitability, uniquely comprising elements of slow and fast dynamics thus enabling weak perturbations of the system to result in large-amplitude deviations before recovery, has fueled studies reporting various effects of noise on temporal and spatially extended dynamical systems for over a decade \cite{jung_prl95,ullner_prl03,lindner_pr04}. It is the different noise dependencies of the slow and fast dynamics, constituting a so-called threshold duality, that are responsible for the majority of noise-induced phenomena observed in excitable systems \cite{pradines_pre99}. In particular, while the slow dynamics, representing the lower threshold, is very susceptible even to weak noisy perturbations, the fast dynamics, representing the upper threshold, is not. On the other hand however, the fast dynamics, resulting in large amplitude excitations, is sensitive to strong noise. Consequently, the temporal order of these excitations exhibits a maximum when we gradually increase the strength of noise \cite{pikovsky_prl97}.

Although some correlations between the coherence resonance in dynamical systems and noise-guided evolution have recently been established in terms of proximity to special bifurcation points \cite{perc_njp06c}, the search for additional conceptual similarities linking the two avenues of research is still vibrant. This paper links these avenues by showing that a reticulate six-species predator-prey model with heterogeneous invasion probabilities offers adjustment possibilities via fine-tuning of a single parameter that takes the system from a noise frigid to a very susceptible state. In the latter state, the studied predator-prey model incorporates two thresholds that can be affected by noise, whereby one is small, conceptually similar to the slow dynamics, and the other large, conceptually similar to the fast dynamics of an excitable system. By introducing Gaussian noise, we show that this threshold duality can be readily exploited in a resonant manner, thus indicating conceptual similarities between the presently reported phenomenon of noise-guided evolution within cyclical interactions and that of coherence resonance reported previously for excitable dynamical systems. We emphasize, however, that the outlined similarities are indeed only conceptual, as the quantity presently of interest is not temporal order of the dynamics as by the observation of coherence resonance but the survival chance of different species within the habitat, which resonantly depends on the intensity of noise. Moreover, the threshold duality is not explicitly constituted by the slow and fast dynamics as by excitable systems, but as we will show below, is routed in the parameters defining the predator-prey interactions.

The remainder of this paper is organized as follows. In the next Section we define the microscopic model and summarize the results for the noise-free case. We also describe the introduced Gaussian noise and comment on its impact on the elementary process. The details of Monte Carlo ($MC$) simulation are provided as well. In Sec.~3 we present the effects of Gaussian noise, and show that at appropriate system conditions the latter can have a large impact on the survival of different species, even reverting the evolutionary process in a resonant manner. In particular, while small and large intensities of noise fail to have a profound impact on the model, intermediate intensities induce a re-entrance effect of the seemingly defeated species, hence decisively guiding the evolutionary process. The non-monotonous effect of different noise intensities is supported by the application of the dynamical mean-field ($DMF$) approximation. To test the generality of our findings, we also perform $MC$ simulations of the model on different host lattice topologies. At the end of Sec.~3, we show that the concept of threshold duality, linking the present results with those obtained previously for noise-driven excitable systems, is not limited to predator-prey cyclical interaction but can be applied to models of evolutionary game theory as well. Finally, Sec.~4 features a summary of main results and concludes by suggesting that the discovered mechanism for noise-guided evolution is applicable under different circumstances, and could thus prove useful in various fields of research such as ecology or economy, where evolutionary processes are often subject to unpredictable factors. 

\section{Predator-prey model}  
\begin{figure} 
\begin{center} \includegraphics[width = 6cm]{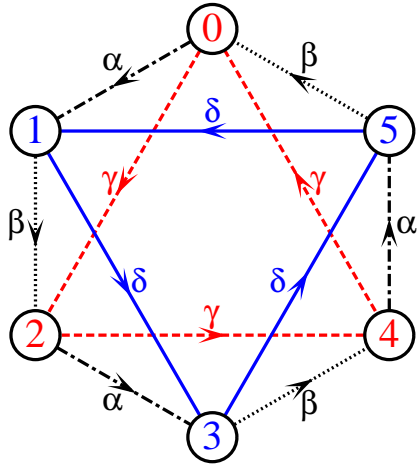} 
\caption{\label{fig:web} Food web of the studied predator-prey model.  Arrows point from predators towards prey with heterogeneous invasion probabilities specified along the edges. Inner loops are colored for easier terminology.} 
\end{center} 
\end{figure}

The studied predator-prey model comprises six species that are initially uniformly distributed on the square lattice. The distribution function of species is given by a set of variables $s_i = 0,\dots, 5$, where $i$ runs over all $L \times L$ lattice sites. Predator-prey relations of nearest neighbors and the corresponding invasion probabilities ($0 < \alpha, \beta, \gamma, \delta < 1$) are defined by the food web presented in Fig.~\ref{fig:web}. More precisely, $\alpha$ determines the probability that an even labeled predator will invade an odd labeled prey, and vice versa for $\beta$, while $\gamma$ and $\delta$ determine probabilities of invasions within even and odd species, respectively. As exemplified in \cite{vikas_pre02,provata_pre03} for lattice Lotka-Volterra type models, the studied six-species model can be defined by a concise reaction scheme. In particular, for species $s = 2k$  
\begin{eqnarray} s \,+ \,j \quad & &\mathop{\rightarrow}^{\alpha} \quad 2 \,s \\ s \,+ \,l \quad & &\mathop{\rightarrow}^{\gamma} \quad 2 \,s  
\end{eqnarray} and for species $s = 2k + 1$ 
\begin{eqnarray} s \,+ \,j \quad & &\mathop{\rightarrow}^{\beta} \quad 2 \,s \\ s \,+ \,l \quad & &\mathop{\rightarrow}^{\delta} \quad 2 \,s  
\end{eqnarray} 
where $j = (s + 1) \,\,$mod $\,\,6$, $l = (s + 2) \,\,$mod $\,\,6$, and $k = 0,1,2$. In case of homogeneous invasion probabilities ($\alpha=\beta=\gamma=\delta=1$) the system has two equivalent three-species states [denoted by $(0+2+4)$ and $(1+3+5)$] exhibiting a self-organizing pattern maintained by cyclic invasions. Departing from a randomly distributed state where every lattice site is occupied by one of the six species, a spatial organization of growing domains begins. The domains are formed either by the members of the red alliance $(0+2+4)$ or the members of the blue alliance $(1+3+5)$ (see Fig.~\ref{fig:web} for the color assignment). In accordance with the coarsening process on the spatial grid, one of the two domains is eliminated with equal probability, and the system ends up in the absorbing state consisting of members of either the red or the blue alliance only. In order to characterize the stationary state more precisely, we may introduce the order parameter $m = \rho_1 + \rho_3 + \rho_5 - \rho_0 - \rho_2 - \rho_4$, whereby  $\rho_s$ ($s=0, \ldots, 5$) denotes the fraction of species $s$ on the spatial grid. Here $m = 1$ corresponds to the exclusive presence of the blue alliance $(1+3+5)$, while $m = -1$ indicates the absolute authority of the red alliance $(0+2+4)$. The two domains are more precisely called "defensive alliances" because their members protect each other cyclically against the external invaders \cite{szabo_pre01b,szabo_jpa05}. For example, if species $0$ invades the allies $(1+3+5)$, in particular by attacking species $1$, its intention is immediately disabled by the species $5$ that is superior to both $0$ and $1$. Thus, the intruder $0$ is quickly abolished from the $(1+3+5)$ domain by the very same species $5$ that dominates species $1$ within the alliance. This reasoning applies for all other possible attempts of non-allied species to invade a defensive alliance. Noteworthy, in this mechanism the proper spatio-temporal distribution of species plays a crucial role, meaning that the classical mean-field approximation, assuming a well-mixed state, cannot reproduce this feature. An interesting possibility is now to study effects of inhomogeneous invasion probabilities, which might affect the evolutionary process in a non-trivial way \cite{frean_prsb01}, essentially forming the basis for noise-guided evolution to be reported below.  

\begin{figure} 
\begin{center} \includegraphics[width = 10cm]{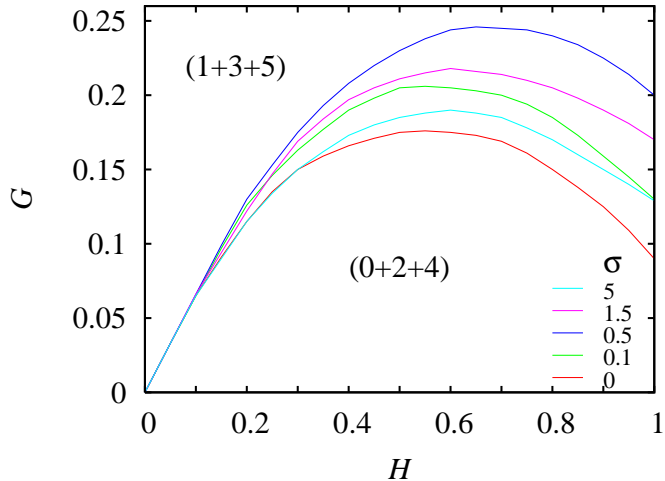} 
\caption{\label{fig:phd} Phase diagrams in dependence on the two main system parameters $H$ and $G$ for various $\sigma$. Lines denote the border separating the two pure phases that are characterized either by the exclusive dominance of the red $(0+2+4)$ ($m = -1$) or the blue $(1+3+5)$ ($m = 1$) alliance.} 
\end{center} 
\end{figure}

Since effects of inhomogeneous invasion probabilities on the evolution of defensive alliances in the studied cyclically dominated model have already been presented in \cite{perc_pre07x}, we here just summarize the main findings that are essential for the present study, and refer the reader to the previous work for more details. First, it is important to note that heterogeneities in the invasion probabilities are introduced in an alliance-specific way. In particular, we can study what happens when one defensive alliance is more aggressive towards the other ($\alpha \neq \beta$), or when the internal mechanism fails to assure flawless protection against the invaders ($\gamma \neq \delta$). In order to address these two issues systematically, we introduce two control parameters that, due to symmetries in the food web, uniquely determine the stationary state of the system characterized by the order parameter $m$. Namely, let $G = \beta - \alpha$ and $H = \gamma - \delta$ where $H, G \in [-1, 1]$. Since $G < 0$ ($\beta < \alpha$) and $H > 0$ ($\gamma > \delta$) clearly favor the survival of the red alliance $(0+2+4)$ (and vice versa for the blue alliance), an interesting competition obviously emerges only if $G > 0$ and $H > 0$ (or equivalently if $G < 0$ and $H < 0$). Due to the symmetry of the problem we restrain or study to the parameter space spanning over $H, G \in [0, 1]$. As soon as $G$ rises above zero (keeping $H = 0$) allies $(1+3+5)$ are favored as their members invade non-allied species more successfully ($\alpha < \beta$), thus $MC$ simulations yielding $m = 1$ in the stationary state. However, the advantage of the blue alliance $(1+3+5)$, given by $G > 0$, may be compensated by choosing sufficiently large values of $H$. Especially, if $H$ rises above zero the internal invasions within the alliance $(1+3+5)$ slow down in comparison to $(0+2+4)$, thus decreasing the effectiveness of the protection shield of the blue alliance and in turn nullifying its advantage given by $G > 0$. Surprisingly the results of $MC$ simulations reveal a non-monotonous phase diagram in dependence on $H$, as shown in Fig.~\ref{fig:phd}. In particular, the advantage of allies $(1+3+5)$ again increases if $H$ approaches 1 (the internal invasion probability $\delta$ vanishes). Note that as $\delta \to 0$ allies $(1+3+5)$ essentially stop to invade each other within the alliance. Via a stability analysis of the interface separating members of the two defensive alliances we have shown \cite{perc_pre07x} that the penetration of the red alliance into the unfavorable $G >0$ region is a consequence of an interface driven effect. Presently, we will show that the spatiality of the system is also a relevant feature when the system is driven from a noise frigid to a noise susceptible state by varying the parameter $H$. 

Turning to the effects of noise, we should point it out that the model involves stochasticity already in its present form. In particular, the invasions of predators to neighboring prey sites are not deterministic but characterized by a probability. The latter source of stochasticity is routed in the spatiality of the model, dictating a random selection of one neighbor of each species in order to carry out an elementary invasion process. Stochastic effects also cannot be neglected if the system size is finite. However, this is a plausible assumption for populations studied within the context of evolutionary game theory \cite{traulsen_jtb07a,traulsen_jtb07b,wild_jtb07}. In these systems the strategy adoption is also ruled by a probability function that depends on the payoff differences between the two competing strategies \cite{szabo_cm06}. Motivated by bounded rationality, often an additional source of stochasticity is introduced by the so-called smooth imitation rule that allows an inferior strategy to replace a more successful one. The application of the Fermi function, adopted from statistical physics, allows to measure the intensity of selection via a single parameter $K$, termed accordingly as the temperature of selection. In the $K \rightarrow 0$ limit the stochastic effects can be neglected, while for $K \rightarrow \infty$ limit the stochasticity is maximal. In our predator-prey model, lacking individual payoffs, the stochasticity resulting from finite $K$ can be modeled by introducing Gaussian noise additively to all invasion probabilities so that the modified invasion probability is $p = x + \xi$ where $x \in (\alpha, \beta, \gamma, \delta)$ and $\xi$ is a random variable being normally distributed with zero mean and $\sigma$ standard deviation. It is worth noticing that the introduced Gaussian noise is in fact multiplicatively coupled with the dynamics of $\rho_s$, as can be inferred from the classical mean-field equations or the $DMF$ approximation. The Gaussian noise satisfies the correlation function  
\begin{equation}
\langle \xi_m (h) \xi_n (k) \rangle = \sigma^2 \delta_{mn} \delta_{hk} 
\end{equation} 
where indexes $m$ and $n$ mark the central location of two predator-prey pairs on the lattice, while $h$ and $k$ denote two consecutive pair interactions. Thus, the stochastic disturbances are delta correlated (uncorrelated) in time as well as among predator-prey pairs. Importantly, if an invasion probability becomes negative ($p<0$) due to noise we allow backwards invasions (a prey can occupy a predator's site), while potential values of invasion probabilities above 1 ($\vert p \vert > 1$) are simply treated as sure events. We thus allow noise-induced reversals of the direction of dominance depicted in Fig.~\ref{fig:web}.

Each elementary step of the $MC$ simulation involves two actions. First, two nearest neighbors are chosen at random, and second, if the two neighboring species form a predator-prey pair (species directly connected by an arrow in Fig.~\ref{fig:web}) the prey (or predator depending on $\xi$) is killed with the probability $p = x + \xi$, where $x$ is the invasion probability specified along the connecting arrow and $\xi$ is the Gaussian noise with properties as described above. On the other hand, if the two randomly chosen species form a neutral pair (species not directly connected by an arrow in Fig.~\ref{fig:web}), or if both are identical, the second part of the elementary $MC$ step dictates no action. In accordance with the random sequential update, each individual is selected once on average during a particular $MC$ step.

\section{Results}
First, we study how the phase separation line in the $H - G$ parameter space varies in dependence on $\sigma$. Due to symmetries in the food web, and to ensure transparency, we perform all below calculations using $\beta = 1.0$ and  $\gamma = 1.0$, thus designating $\alpha$ and $\delta$ as the only parameters able to change $H$ and $G$. Figure~\ref{fig:phd} shows the results, from which it can be inferred that there exist an intermediate $\sigma$ for which allies $(0+2+4)$ receive the biggest boost, or in other words, are able to compensate for the largest values of $G$ by a given $H$. Clearly, there exist a resonant dependence on $\sigma$ as larger levels of stochasticity fail to have the same effect because the phase separation line gradually converges to the zero-noise limit as $\sigma$ increases.

\begin{figure} 
\begin{center}
\includegraphics[width = 10cm]{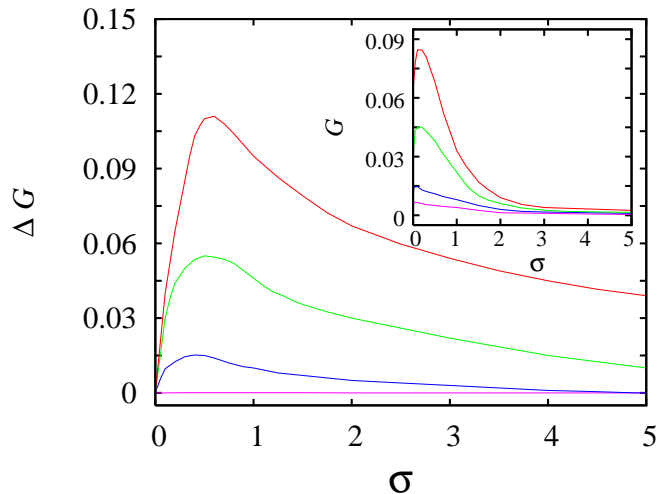} 
\caption{\label{fig:res} Absolute rise of the phase separation line $\Delta G$ in dependence on $\sigma$ for different values of $H$ (pink$ = 0.1$, blue$ = 0.2$, green$ = 0.5$, red$ = 0.9$). Note that positive values of $\Delta G$ manifest a facilitative effect of noise on the survival of the red alliance $(0+2+4)$. The inset shows results from the four-point $DMF$ approximation for the same values of $H$ as in the main figure, in particular depicting critical values of $G$ separating the two pure phases.} 
\end{center} 
\end{figure}

To study the outlined resonant dependence, we plot $\Delta G$ as a function of $\sigma$, whereby $\Delta G$ measures the rise of the phase separation line in the vertical direction with respect to its position at $\sigma = 0$ at any given $H$. Results for different values of $H$ are presented in Fig.~\ref{fig:res}. Two important conclusions can be made. First, added Gaussian disturbances affect the evolutionary process resonantly in dependence on $\sigma$, and second, the phenomenon becomes increasingly pronounced as $H \to 1$ ($\delta \to 0$). We should also point out that by values of $H$ close to 0 noise has absolutely no impact on the evolutionary process. Thus, we demonstrate that as the parameter $H$ is varied from 0 to 1 the system goes from a noise frigid to a noise susceptible state, and moreover, noise-guided evolution sets in if $\sigma$ is appropriately adjusted.

Our numerical findings can be supported by an alternative approach entailing the application of $k$-site cluster $DMF$ approximation. Since the classical mean-field approximation assumes well-mixed populations, it is easy to understand that it cannot reproduce above-described results. As we have argued earlier, the spatiality plays a fundamental role by the explanation of the penetration of the red alliance into the unfavorable $G>0$ area. The shortage of the classical mean-field approximation may be eliminated by applying the $DMF$ approximation technique that proved to be very appropriate for obtaining qualitatively correct phase diagrams for several non-equilibrium systems. For a detailed description of this method we refer the reader to earlier papers \cite{ szabo_cm06,dickman_pre01}. The $DMF$ approximation is a dynamical version of the cluster-variational method, and it involves finding a hierarchy of evolution equations for the probability distributions of configurations within a cluster of $k$ sites. If correlations in larger clusters are neglected then the level of approximation can be characterized by the value of $k$. Similarly as by the $MC$ simulations, in the $DMF$ approach the introduction of Gaussian noise can, besides altering the strength of invasions, also revert the direction of dominance. Therefore, aside from handling the elementary processes displayed in Fig.~\ref{fig:web}, we also have to consider the possibility of reverse processes. With this extension, the four-point level of approximation can describe correctly not only the impact of the dynamical benefit of the red alliance, but also the impact of Gaussian noise. Indeed, displayed as an inset in Fig.~\ref{fig:res}, the re-entrance phenomenon driven by increasing values of $\sigma$ is well reproduced, and moreover, exhibits the same dependence on $H$ as revealed by $MC$ simulations. In particular, values of $H$ close to zero yield a virtually noise-resistant dynamics across the whole span of $\sigma$, whereas noise-guided evolution sets in as $H \to 1$ for intermediate levels of stochasticity. Nonetheless, a slight difference between the results of $MC$ simulations and the $DMF$ approximation can be observed. According to the latter, large values of $\sigma$ can eliminate the dynamical benefit of allies $(0+2+4)$, thus resulting in the solution predicted by the classical mean-field theory. An even higher level of approximation might still eliminate this minor inconsistency with the $MC$ simulations, but unfortunately the resulting number of variables is too large to be manageable by the computer resources currently available to us. In sum, the failure of classical noise-driven mean-field equations to reproduce the results of $MC$ simulations and the necessity for the four-point cluster $DMF$ approximation confirm that the reported noise-guided evolution within cyclical interactions is indeed deeply routed in the short-range correlations of the spatial distribution, as conjectured already from the outlay of phase diagrams depicted in Fig.~\ref{fig:phd} and findings reported previously in \cite{perc_pre07x}.

\begin{figure} 
\begin{center} \includegraphics[width = 10cm]{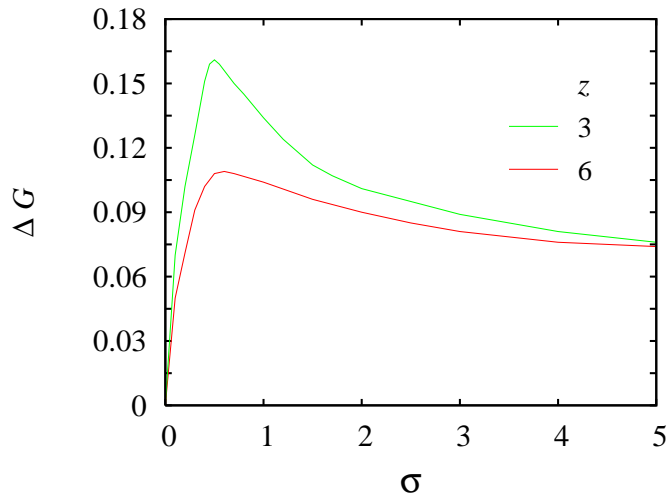} 
\caption{\label{fig:gr} Absolute rise of the phase separation line $\Delta G$ in dependence on $\sigma$ for the honeycomb (green) and the triangular (red) lattice, obtained by $H = 0.99$. The interpretation of $\Delta G$ is the same as by Fig.~\ref{fig:res}. Importantly, the attainable effect of noise decreases as the coordination number $z$ increases.} 
\end{center} 
\end{figure}

In order to strengthen our arguments, we also examine the dynamics of the noise-driven model via $MC$ simulations on two additional types of regular graphs, namely on the honeycomb and the triangular lattice. Our choice is motivated by the fact that the two types of host lattices have different coordination number $z$. While the honeycomb lattice ($z = 3$) allows manifesting the importance of spatial correlations, the triangular lattice ($z = 6$) may move the system closer towards mean-field conditions. Therefore we argue that the noise driven re-entrance phenomenon will be more (less) pronounced when the honeycomb (triangular) host topology is applied. Indeed, results presented in Fig.~\ref{fig:gr} fully support this expectation as the observed phenomenon is substantially better expressed for $z = 3$ than for $z = 6$, and thus additionally strengthen the fact that the reported noise-guided evolution within cyclical interactions is routed in the short-range correlations of the spatial distribution.   

Next, it is of interest to draw some analogies between the reported noise-guided evolution within cyclical interactions and the phenomenon of coherence resonance reported previously in excitable dynamical systems \cite{pikovsky_prl97,pradines_pre99}. Although we presently consider the survival of different alliances as being resonantly dependent on $\sigma$ and not the temporal order of their evolution, and are thus strictly speaking unable to write about a classical coherence resonance phenomenon, we still argue a conceptual link can be established by addressing the presently reported phenomenon as an evolutionary coherence resonance \cite{perc_njp06c}. Noteworthy, related conceptual differences emerge also in \cite{traulsen_prl04}, where the authors employs similar arguments to address the problem. Aside from this difference, however, we will show that the threshold duality, constituted by the noise susceptible slow dynamics and the noise robust fast dynamics of excitable systems, can be identified also in the present model, and is in fact crucial for the observation of noise-guided evolution.

The threshold duality can be made visible by studying the properties of invasion probabilities along the phase separation line of the deterministic model shown in Fig.~\ref{fig:phd}. We start with close to zero values of $H$ where, taking into account results presented in Fig.~\ref{fig:res}, noise has no impact on the evolutionary process irrespective of $\sigma$. There, the switch between the dominance of allies $(1+3+5)$ and $(0+2+4)$ occurs at $H = 0.1$ and $G = 0.065$, or equivalently, at $\delta = 0.9$ and $\alpha = 0.935$ (note that $\beta = 1.0$ and $\gamma = 1.0$ are held fixed throughout this work). Importantly, the differences $\delta_{th} = 1 - \delta$ and $\alpha_{th} = 1 - \alpha$ define the two thresholds Gaussian noise can influence in order to guide the evolutionary process. In particular, if one of the two invasion probabilities $\alpha$ or $\delta$ would become equal to 1, the corresponding defensive alliance would benefit and eventually defeat the other, as can be inferred from the read line depicted in Fig.~\ref{fig:phd} (consider for example the case when $\delta = 1.0$ while $\alpha = 0.935$; clearly the blue alliance wins making $m = 1$ in the stationary state). However, since at $H = 0.1$ both thresholds have virtually identical magnitude ($\delta_{th} = 0.1$ and $\alpha_{th} = 0.065$), and are exposed to the same intensity of temporally and spatially uncorrelated Gaussian noise, the long-term probability of exceeding one threshold while leaving the other untouched is practically zero. Thus, noise cannot influence the evolution of the two defensive alliances irrespective of $\sigma$, as displayed in Fig.~\ref{fig:res}. However, the situation changes substantially as $H$ increases. By $H = 0.5$ the phase separation line is situated at $G = 0.175$ in the noise-free case. The two relevant thresholds then equal $\delta_{th} = 0.5$ and $\alpha_{th} = 0.175$. As above, if Gaussian noise would exceed only $\alpha_{th}$ the red alliance would win ($m = -1$), and via similar reasoning, if only $\delta_{th}$ would be exceeded the blue alliance would win ($m = 1$). Contrary to the $H = 0.1$ case here the two relevant thresholds differ substantially. Therefore it is easy to see that by an appropriate value of $\sigma$ the threshold $\alpha_{th} = 0.175$ will be exceeded statistically far more often than $\delta_{th} = 0.5$, which ultimately results in noise-guided evolution, specifically giving the seemingly defeated alliance $(0+2+4)$ the winning edge over the allies $(1+3+5)$, as indicated by the positive values of $\Delta G$ in Fig.~\ref{fig:res}. However, by larger values of $\sigma$ the upper threshold $\delta_{th} = 0.5$ will be exceeded often as well, thus again nullifying the dynamical benefit coming from the faster inner cycle for the red alliance, which explains the depicted resonant dependence of $\Delta G$. Note that although larger $\sigma$ increase the probability of crossing the lower threshold ($\alpha_{th} = 0.175$) as well, the maximal value of each invasion probability equaling 1 (above that each event is simply treated as a sure event) prohibits a noticeable effect of this fact, and thus results in the decline of $\Delta G$ as $\sigma$ increases. We argue that the difference in the two thresholds (lower threshold given by $\alpha_{th} = 0.175$ and the upper one given by $\delta_{th} = 0.5$), constituting a threshold duality in the presently studied model of cyclical interactions, is conceptually similar to the threshold duality observed in excitable dynamical systems, albeit the latter is routed in the fast and slow dynamics of the system \cite{pikovsky_prl97,pradines_pre99}, while the former originates directly from the parameters values defining the evolutionary process.

\begin{figure} 
\begin{center} \includegraphics[width = 10cm]{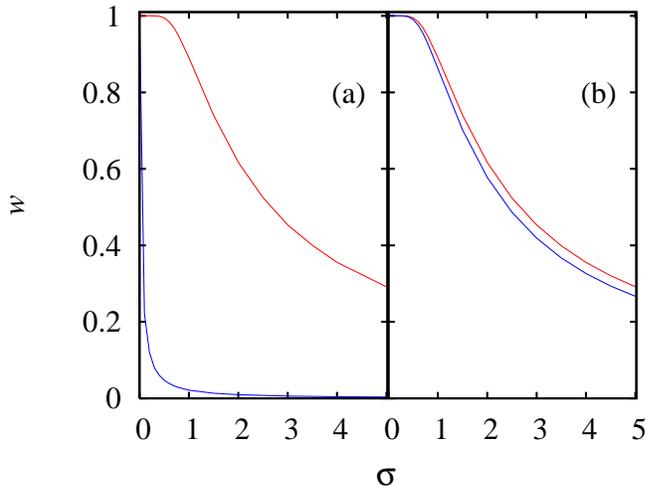} 
\caption{\label{fig:weak} Normalized difference of clockwise and anti-clockwise invasions $w$ along the two inner circles (depicted with blue and red lines in accordance with the coloring in Fig.~\ref{fig:web}) of the food web in dependence on $\sigma$. Panel (a) show results for $H = 0.99$ and $G = 0.07$, while panel (b) features results for $H = 0.1$ and $G = 0.09$.} 
\end{center} 
\end{figure}

Finally, it is straightforward to extend the above argument also to the case when $H = 0.99$, whereby the noise-free phase separation line is situated at $G = 0.09$, and hence the two relevant thresholds, equaling $\delta_{th} = 0.99$ and $\alpha_{th} = 0.09$, clearly differ by an order of magnitude thus constituting the vital threshold duality needed for the observation of noise-guided evolution. Importantly, an additional threshold emerges related to the possibility of noise-induced negative values of $\delta$. Specifically, the new threshold simply equals to $\delta = 0.01$, and is apparently of similar magnitude as $\alpha_{th} = 0.09$. Since negative values of $\delta$ flip the circle of dominance within the blue alliance $(1+3+5)$, thus temporarily disrupting its protective shield against the invaders, intermediate values of $\sigma$ warrant a two-fold advantage to the red alliance, ultimately resulting in the best-expressed resonance curve in Fig.~\ref{fig:res}. The origin of the two-fold advantage of the red allies can be studied more precisely by measuring the success rate of invasions along the two inner circles (depicted with blue and red in Fig.~\ref{fig:web}) of the food web. To be more specific, we calculate the normalized difference of clockwise and anti-clockwise invasions $w$ for the red and blue alliance in dependence on $\sigma$. Note that $w = 1$ if all invasions occur in the anti-clockwise direction (as depicted by arrows in Fig.~\ref{fig:web}) and zero if clockwise and anti-clockwise invasions occur equally often. In other words, the value of $w$ characterizes the strength of the defensive mechanism within an alliance. Results in Fig.~\ref{fig:weak} show that the defensive shield of the blue alliance $(1+3+5)$ suffers virtually immediately after noise is introduced, whereas the red alliance is able to uphold a perfectly functioning protection until $\sigma \approx  0.6$. Naturally, as $\sigma$ increases further the defensive capacity of the red alliance also weakens. However, the window resulting from the delayed faint of the defensive mechanism, in conjunction with the previously described mechanism warranted by the threshold duality via $\delta_{th}$ and $\alpha_{th}$, enables a resonant strength of the referred $(0+2+4)$ domain. Results displayed in Fig.~\ref{fig:res} support this argument since $\sigma \approx 0.6$ limits the peak of the resonance curve obtained for $H = 0.99$, thus signaling the beginning of end of the noisy support for the red allies. Also in accordance with the results presented in Fig.~\ref{fig:res}, for $H = 0.1$ the dependence of $w$ on $\sigma$ is very similar for both defensive alliances (red and blue), thus clearly signaling the absence of the mechanism warranting noise-guided evolution in the studied model. The simultaneous decrease of the defensive strength results in a noise-frigid state since all invasion probabilities are of similar magnitude. The above-described simple arguments provide a workable mechanism for noise-guided evolution, and at least conceptually unify the phenomenon with the coherence resonance in excitable dynamical systems. 

To end this section, we would like to note that the above explanation warranting the resonant dependence of  $\Delta G$ on $\sigma$, intimately relying on the existence of threshold duality in evolutionary processes, is valid for other systems as well. To demonstrate this, we briefly consider a spatial prisoner's dilemma game \cite{nowak_n92b,nowak_ijbc94,hauert_ajp05}, where two cooperators receive the reward $R = 1$, two defectors receive the punishment $P = 0$, whilst a cooperator and defector receive the suckers payoff $S = -r$ and the temptation $T = 1 + r$, respectively, thus satisfying the prisoner's dilemma payoff ranking $T > R > P > S$ if $r > 0$. It is easy to identify a threshold duality in the system if only $r \ll 1$. If the payoffs become stochastic as a result of additive Gaussian noise \cite{perc_pre07}, inequalities $T > R$ and $P > S$ differing by $r$ will be violated statistically much more frequently by a given $\sigma$ than the $R > P$ inequality differing by 1. It is easy to understand that intermediate levels of stochasticity promote cooperation because the violation of $T > R$ and $P > S$ favors the cooperative strategy since it potentially nullifies the advantage $r$ defectors have over cooperators (two cooperators might end up receiving a larger payoff each than a defector facing a cooperator, and a cooperator facing a defector might be better off than two defectors). On the other hand, this facilitative effect is limited by violations of $R > P$, belonging to the so-called social dilemma \cite{macy_pnas02,santos_pnas06}, because then two defectors might be better off than two cooperators, which again gives the winning edge to the defecting strategy, and hence results in a resonant dependence of cooperation fitness on $\sigma$ \cite{perc_njp06a}. Although being fairly simple, the described explanation outlines a general mechanism of cooperation promotion in the spatial prisoner's dilemma game, thus giving another example of noise-guided evolution relying on the threshold duality in evolutionary processes. 

\section{Summary}
In sum, we provide conclusive evidences that the threshold duality is crucial for the observation of noise-induced resonances in evolutionary processes. Particularly, we show that a reticulate six-species predator-prey model with heterogeneous invasion probabilities possesses a single parameter that is able to guide the system from a noise frigid to a noise susceptible state, thereby relying exclusively on the emergence of threshold duality. The phenomenon revealed by $MC$ simulations can be verified by a four-point cluster $DMF$ approximation, thus providing ample evidences that the reported noise-driven evolution within cyclical interactions is heavily routed in the short-range correlation of the spatial distribution. This conjecture is additionally supported by studying the evolution of species on different regular graphs, and confirmed by the fact that the phenomenon of noise-guided evolution weakens as the coordination number of the host lattice increases. While the observations in the studied model are related to the system-specific non-monotonous phase separation line between the two defensive alliances, we note that the general concept is valid also in the framework of evolutionary game theory, thus suggesting it may be widely applicable in various fields of research ranging from economy to ecology.
\ack
M. P. acknowledges support from the Slovenian Research Agency (grant Z1-9629). Research of A. S. was supported by the Hungarian National Research Fund (T-47003, NKTH-MFAIIF05).

\end{document}